\documentclass[aps,prc,preprintnumbers,superscriptaddress,floatfix,showpacs, twocolumn]{revtex4-1}

\usepackage{epsfig}
\usepackage{amsmath}

\begin{document}
\title{Hydrodynamic Predictions for Mixed Harmonic Correlations in 200 GeV Au+Au Collisions}
\author{Fernando G. Gardim}
\affiliation{Instituto de Ci\^encia e Tecnologia, Universidade Federal de Alfenas, Cidade Universit\'aria, 37715-400 Po\c cos de Caldas, MG, Brazil}
\author{Frederique Grassi}
\affiliation{Instituto de F\'{i}sica, Universidade de S\~{a}o Paulo, 05315-970 S\~{a}o Paulo, SP, Brazil}
\author{Matthew Luzum}
\affiliation{Instituto de F\'{i}sica, Universidade de S\~{a}o Paulo, 05315-970 S\~{a}o Paulo, SP, Brazil}
\author{Jacquelyn Noronha-Hostler}
\affiliation{Department of Physics, University of Houston, Houston, TX 77204, USA}
\begin{abstract}
Recent measurements at the LHC involve the correlation of different azimuthal flow harmonics $v_n$.  These new observables add constraints to theoretical models and probe aspects of the system that are independent of the traditional single-harmonic measurements such as 2- and multi-particle cumulants $v_n\{m\}$.  Many of these new observables have not yet been measured at RHIC, leaving an opportunity to make predictions as a test of models across energies. 
We make predictions using NeXSPheRIO, a hydrodynamical model which has  accurately reproduced a large set of single-harmonic correlations in a large range of transverse momenta and centralities at RHIC.  Our predictions thus provide an important baseline for comparison to correlations of flow harmonics, which contain non-trivial information about the initial state as well as QGP transport properties.  We also point out significant biases that can appear when using wide centrality bins and non-trivial event weighting, necessitating care in performing experimental analyses and in comparing theoretical calculations to these measurements.
\end{abstract}
\maketitle
\section{Introduction}
In the standard picture of a heavy-ion collision, after a short period of non-equilibrium dynamics, the system expands as a relativistic fluid.  As the system cools, the Quark Gluon Plasma transitions into hadrons that eventually spread apart sufficiently so  interactions cease, after which the particles are detected.

In theory, one can understand these particles as having been emitted according to an underlying distribution, which can be written as a Fourier series with respect to the azimuthal angle $\phi$
\begin{align}
P(\phi) &= \frac 1 {2\pi} \sum_n V_n e^{-in\phi} \\
&= \frac 1 {2\pi} \sum_n v_n e^{i\Psi_n}e^{-in\phi} .
\end{align}
The parameters $v_n$ and $\Psi_n$ are the magnitude and orientation of the flow vectors (written compactly as the magnitude and phase of a complex number $V_n$), each of which can depend on the other two degrees of freedom --- transverse momentum and pseudorapidity.   These flow vectors fluctuate significantly from one event to the next, even within a particular centrality class. Thus, there is not a small set of constant coefficients $\{v_n, \Psi_n\}$, but instead a large set of statistical properties.  

For example, the magnitude squared of each flow vector, $|V_n|^2 = v_n^2$, has an entire event-by-event distribution that has been measured and analyzed in \cite{Aad:2013xma}.  Similarly, the correlation between flow vectors at different points in pseudorapidity and transverse momentum has been a recent topic of interest \cite{Gardim:2012im, Bozek:2010vz, Xiao:2012uw, Jia:2014ysa, Khachatryan:2015oea,Jia:2012ez,Jia:2014vja}. Recent studies have shown the neccessity of event-by-event fluctuations at high transverse momentum as well \cite{Noronha-Hostler:2016eow}.  

However, there is much more information present in the event-by-event distribution of particles.  The mentioned quantities involve only a single Fourier harmonic $n$.  One can also consider the alignment and correlation between flow vectors of different harmonics, which opens the door to a large number of additional measurements.  Many of these have now been performed by LHC collaborations \cite{Aad:2014fla, ALICE:2016kpq,Jia:2014jca}, providing new constraints on theory, and yet others have been suggested \cite{Yan:2015jma,Bhalerao:2014xra}.   Most have not yet been done at RHIC, leaving an opportunity to make predictions.

Similarly to how adding measurements of $v_3$ in addition to $v_2$ provided significant constraints on, e.g., the initial stages of the collision \cite{Adare:2011tg}, these new mixed-harmonic correlations provide non-trivial constraints on theory.  Models that fit traditional observables well can not necessarily fit these new observables \cite{Niemi:2015qia,Zhou:2015eya}, which thus provide added insight into both initial conditions and medium properties.  
For example, combinations of these observables may be able to isolate linear and non-linear hydrodynamic response  \cite{Giacalone:2016afq} or shed light on the temperature dependence of $\eta/s$  \cite{Niemi:2015qia}.

Our main result is a prediction for these upcoming measurements at RHIC. In addition, we point out that details of experimental analyses such as centrality binning and event weighting can have important effects, which must be taken into account for apples-to-apples comparisons of theory to experiment.   In Appendix \ref{anc}, we also make some observations about the dependence on viscosity, collision energy, and initial conditions.  
\section{Mixed Harmonic Correlations}
The basic building block of correlation measurements is the general $m-$particle correlator \cite{Bhalerao:2011yg, Luzum:2013yya, Bilandzic:2013kga}
\begin{widetext}
\begin{align}
\label{corr}
\langle m \rangle_{n_1, n_2, \ldots, n_m}
&\equiv \biggr\langle \langle \cos(n_1 \phi^{a_1} + n_2 \phi^{a_2}\ldots + n_m \phi^{a_m}) \rangle_{m\ \rm particles}\biggl\rangle\\
\label{corrflow}
&\stackrel{\rm{(flow)}}{=} \left\langle V_{n_1}^{a_1} V_{n_2}^{a_2} \ldots V_{n_m}^{a_m} \right\rangle\\
&\stackrel{\rm{(flow)}}{=} \left\langle v_{|n_1|}^{a_1} v_{|n_2|}^{a_2} \ldots v_{|n_m|}^{a_m} \cos(n_1\Psi_{n_1}^{a_1} + n_2\Psi_{n_2}^{a_2}\ldots n_m \Psi_{n_m}^{a_m}) \right\rangle  \;.
\label{corrflow2}
\end{align}
\end{widetext}
Here the inner average is over all possible groupings of $m$ particles and the outer average is over all events. 
Rather than simple averages, these can be weighted averages.  For example, particles can be weighted by their transverse momentum or pseudorapidity.  Or, more relevant to measurements considered in this work, one can weight each event based on the number of charged hadrons in that event.  Explicitly:

\begin{equation}
\label{weight}
\left\langle \ldots \right\rangle \equiv \frac {\sum_{\rm events} W \ldots} {\sum_{\rm events} W}
\end{equation}

A simple average has $W$ = 1.  Another possible choice \cite{Bilandzic:2010jr} is to use the number of combinations $W = M!/(M-m)! \simeq M^m$, where $m$ is the number of particles in the correlation \eqref{corr} and $M$ is the total number of particles.  Explicitly, for 2-particle correlations and 4-particle correlations, respectively,
\begin{align}
W_{\langle 2 \rangle} &= M(M-1) \nonumber \\
W_{\langle 4 \rangle} &= M(M-1)(M-2)(M-3)
\label{w24}
\end{align}
 Events with larger multiplicity have a smaller statistical uncertainty.  As a result, biasing your average toward those events reduces the statistical uncertainty in the measurement (this particular choice means each pair or quadruplet of particles has equal contribution, rather than each event).   As we will show, this bias can have a non-negligible effect, particularly for measurements which are combinations of 2- and 4-particle correlations, so it is important to take this into account when comparing calculations to measurements done with weights.

In Eqs.\ (\ref{corr}--\ref{corrflow2}) the labels $a_i$ represent bins in momentum and/or particle species, from which the $i$th particle is chosen.  In principle, the momentum bin of each particle can be varied independently, and the correlator is therefore a function of $2m$ degrees of freedom --- the transverse momentum and pseudorapidity of each particle.  Because each collision has a random azimuthal orientation, one can only measure rotation-invariant quantities.  Therefore, the analogous correlations involving sine instead of cosine vanish, and non-zero correlators must have $\sum n_i = 0$.    In principle, this set of correlators contains all available information about particle correlations. 

In a pure hydrodynamic calculation (as well as a description purely in terms of the Boltzmann equation) all particles are uncorrelated, and can be understood as independent samples of the single-particle distribution, which then contains all possible information about any $m$-particle distribution.  In reality, this may not always be a good approximation, and particles can be correlated by various processes (e.g., resonance decays, jet fragmentation, etc.).  However, it is of significant interest to measure properties of the single-particle distribution, as they contain information about the properties of the quark-gluon plasma as well as certain information about the initial stages of the collision.  Because of this, measurements are often designed in order to suppress any non-flow correlations and isolate statistical properties of the flow vectors $\{v_n, \Psi_n\}$.  

Equations \eqref{corrflow} and \eqref{corrflow2}, then, represent the dependence on the underlying single-particle emission probability, under the assumption that particles are independent.  

Here we concentrate on momentum-integrated measurements, so that all particles are chosen from a wide range of transverse momentum and pseudorapidity (specifically, we will present predictions for $p_T>200$ MeV and $|\eta| < 1$, to match the  acceptance of the STAR TPC).

The simplest and most common measurements are two-particle cumulants, also known as the scalar product $v_n$ \cite{Adler:2002pu}
\begin{align}
v_n\{2\} = v_n\{SP\} &= \sqrt{\langle 2 \rangle_{n,-n}} \\
& \stackrel{\rm{(flow)}}{=} \sqrt{ \langle v_n^2\rangle} ,
\end{align}
which measures the RMS value of $v_n$, in the absence of non-flow correlations.  In general,  non-flow correlations will not necessarily be negligible.  One of the simplest ways to suppress such correlations is to impose a rapidity gap --- that is, to choose only pairs of particles that are widely separated in pseudorapidity.  Most known sources of non-flow occur within a short-range in rapidity, thus, this can be an effective method for selecting flow effects. 

An alternative is to consider correlations between more than 2 particles, in a way that correlations between small numbers of particles are suppressed, and true many-particle correlations can be measured.  The standard measurements of this type are $m$-particle cumulants $v_n\{m\}$ \cite{Borghini:2001vi}, the first few of which are defined as \cite{Bilandzic:2012wva}:

\begin{widetext} 
\begin{align}
\label{vn4}
- v_n\{4\}^4 &\equiv \langle 4 \rangle_{n, n, -n, -n} - 2  \langle 2 \rangle_{n,-n}^2  \\
&\stackrel{\rm{(flow)}}{=} \langle v_n^2\rangle - 2 \langle v_n^2\rangle^2 \\
4v_n\{6\}^6 &\equiv \langle 6\rangle_{n,n,n,-n,-n,-n} - 9 \langle 4 \rangle_{n,n,-n,-n}\langle 2\rangle_{n,-n} + 12 \langle 2 \rangle_{n,-n}^3\\
&\stackrel{\rm{(flow)}}{=} \langle v_n^6 \rangle - 9 \langle v_n^4\rangle \langle v_n^2\rangle + 12 \langle v_n^2 \rangle^3\\
-33v_n\{8\}^8 &\equiv \langle 8\rangle_{n,n,n,n,-n,-n,-n,-n} - 16 \langle 6\rangle_{n,n,n,-n,-n,-n}\langle 2\rangle_{n,-n} - 18 \langle 4 \rangle_{n, n, -n, -n}^2\\
&\ \ \ \ \ \ + 144 \langle 4 \rangle_{n,n,-n,-n}\langle 2\rangle_{n,-n}^2 - 144 \rangle_{n,-n}^4 \\
&\stackrel{\rm{(flow)}}{=} \langle v_n^8\rangle - 16  \langle v_n^6 \rangle \langle v_n^2\rangle - 18 \langle v_n^4\rangle^2 + 144 \langle v_n^4\rangle\langle v_n^2\rangle - 144 \langle v_n^2\rangle^4 .
\end{align}
\end{widetext}
Besides suppressing non-flow correlations of order less than $m$, each of these measures a different moment of the event-by-event distribution of the magnitude squared of the flow vector $v_n^2$, and contains independent information about hydrodynamic initial conditions and/or medium properties.

While these observables only contain information about a single Fourier harmonic, $n$, similar measurements can be made involving more than one harmonic.  In the same spirit of the suppression of low-order non-flow correlations, it was proposed to measure mixed harmonic observables based on 4-particle correlations, SC$(n,m)$, or Symmetric Cumulants (originally ``Standard Candles'') \cite{Bilandzic:2013kga},
\begin{align}
{\rm SC}(n,m) &\equiv \langle 4 \rangle_{n,m, -n, -m} - \langle 2 \rangle_{n,-n}\langle 2 \rangle_{m,-m} \\
&\stackrel{\rm{(flow)}}{=} \langle v_n^2 v_m^2 \rangle -  \langle v_n^2\rangle \langle v_m^2 \rangle .
\end{align}
These cumulants are defined only for $n\neq m$, in which case the factor 2 disappears from Eq. \eqref{vn4}.  

Since the information about the overall magnitude of $v_n$ is contained in previous conventional measurements, the independent information can best be viewed with a normalized version of the correlation, 
\begin{align}
\label{NSC}
{\rm NSC}(n,m) &\equiv \frac {{\rm SC}(n,m)} {\langle 2 \rangle_{n,-n}\langle 2 \rangle_{m,-m}} \\
&\stackrel{\rm{(flow)}}{=}  \frac {\langle v_n^2 v_m^2 \rangle -  \langle v_n^2\rangle \langle v_m^2 \rangle} {\langle v_n^2\rangle \langle v_m^2 \rangle} .
\end{align}

These momentum-integrated measurements are only sensitive to the magnitude squared of the flow vector, $v_n^2$.  In order to gain information about the correlations of the entire momentum-integrated flow vector $V_n$, including its direction, one must consider other correlations.   

The ATLAS Collaboration has measured a large set of such correlations \cite{Aad:2014fla}, a few of which are listed as follows
\begin{align}
\label{EP24}
\langle \cos 4\left(\Phi_2 -  \Phi_4 \right) \rangle\{ {\rm SP}\} &\equiv \frac { \langle 3 \rangle_{2,2,-4}} {\sqrt{\langle 4 \rangle_{2,2,-2,-2} \langle 2\rangle_{4,-4}}} \\
&\stackrel{\rm{(flow)}}{=} \frac {\langle V_2^2 V_4^{*} \rangle} {\sqrt{\langle |V_2|^4 \rangle \langle |V_4|^2\rangle}} \\
&\stackrel{\rm{(flow)}}{=} \frac {\langle v_2^2 v_4 \cos 4(\Psi_2 - \Psi_4) \rangle} {\sqrt{\langle v_2^4 \rangle \langle v_4^2\rangle}}
\end{align}
\begin{align}
\label{EP23}
\langle \cos 6\left(\Phi_2 -  \Phi_3 \right) \rangle\{{\rm SP}\} &\equiv \frac { \langle 5 \rangle_{2,2,2,-3, -3}} {\sqrt{\langle 6 \rangle_{2,2,2,-2,-2,-2} \langle 4\rangle_{3,3,-3,-3}}} \\
&\stackrel{\rm{(flow)}}{=} \frac {\langle V_2^3 V_3^{*2} \rangle} {\sqrt{\langle |V_2|^6 \rangle \langle |V_3|^4\rangle}} \\
&\stackrel{\rm{(flow)}}{=} \frac {\langle v_2^3 v_3^2 \cos 6(\Psi_2 - \Psi_3)\rangle} {\sqrt{\langle v_2^6 \rangle \langle v_4^4\rangle}} 
\end{align}
\begin{align}
\label{EP235}
\langle \cos (2\Phi_2 +  3 \Phi_3 & -  5\Phi_5 )  \rangle  \{{\rm SP}\} \equiv \frac { \langle 3 \rangle_{2,3,-5}} {\sqrt{\langle 2 \rangle_{2,-2} \langle 2 \rangle_{3,-3} \langle 2\rangle_{5,-5}}} \\
&\stackrel{\rm{(flow)}}{=} \frac {\langle V_2 V_3 V_5^*\rangle} {\sqrt{\langle |V_2|^2 \rangle \langle |V_3|^2\rangle \langle |V_5|^2 \rangle}} \\
&\stackrel{\rm{(flow)}}{=} \frac {\langle v_2 v_3 v_5 \cos (2\Psi_2 + 3\Psi_3 - 5\Psi_5) \rangle} {\sqrt{\langle v_2^2 \rangle \langle v_3^2\rangle \langle v_5^2 \rangle}}.
\end{align}  
Once again, the lower expressions represent the dependence on hydrodynamic quantities, in the absence of non-flow correlations.  

However, unlike the case of cumulant measurements (both traditional and symmetric), non-flow correlations here are not naturally suppressed, and one instead must maintain a gap in pseudorapidity between particles with a plus sign and particles with a minus sign in both the the numerator and denominator of the correlation definition.  The naive procedure of calculating such correlations with nested loops is computationally prohibitive, and so imposing such a gap on the basis of each group of $m$ particles is unfeasible.  Instead, one uses a ``scalar product'' (SP) procedure, where only one pass through the data is required.  In this case, rapidity gaps are ensured by segmenting the detector, such that each of the $m$ particles comes only from a particular segment.  In this way, the former group of particles can always be separated from the latter.  We note that, while ATLAS used three separate segments to measure ``three-plane'' quantities such as $\langle \cos\left(2\Phi_2 + 3 \Phi_3 - 5\Phi \right) \rangle\{SP\} $, only two are necessary to suppress short-range correlations \cite{Bhalerao:2013ina}, similar to how $\langle \cos 6\left(\Phi_2 -  \Phi_3 \right) \rangle\{SP\}$ can be measured with only two segments, despite involving a 5-particle correlation.

We also note that, despite the original notation chosen by ATLAS, and the name that is often used to describe these measurements (``Event Plane Correlations''), these observables do not in fact measure correlations between event planes $\Psi_n$, but rather the entire flow vector $V_n$, which is not constrained to fluctuate in angle only, as is apparent from the expressions above.  

For a recent review of mixed harmonic correlations, see Ref.~\cite{Zhou:2016eiz}

Finally, we note that the lower-energy collisions at RHIC have a smaller multiplicity, and the STAR and PHENIX detectors have smaller coverage than ATLAS (or CMS).  As a result, the statistical uncertainty on all of these measurements is significantly larger, and not all of the measurements made at the LHC are possible at RHIC.  Because of this, we display here only selected results, in anticipation of those that we expect to be easiest to measure.  However, a much larger set has been calculated and is available.
Note also that there is some arbitrariness in the denominator of the event plane correlations (e.g., using $\langle v_n^4 \rangle$ vs. $\langle v_n^2\rangle^2$).  Results using other definitions are also available.
\section{Calculations}
NeXus initial conditions for Au+Au collisions are generated using a parton-based Gribov-Regge picture of nucleus-nucleus collisions in which hard partons are treated using perturbative QCD while soft partons are included using the string picture.  Details can be found in Ref.~\cite{Drescher:2000ec}. 2000 events are used for each 10\% centrality bin.  

The resulting energy and momentum distribution in each event is then used as an initial condition for 3+1 dimensional ideal hydrodynamic evolution, followed by hadronic decays, using the NeXSPheRIO hydrodynamic code \cite{Hama:2004rr}.  

This setup has been tested extensively and shown to provide a reasonable description of data at top RHIC energy: rapidity and transverse momentum spectra \cite{Qian:2007ff}, elliptic flow \cite{Andrade:2006yh,Andrade:2008xh}, standard directed flow \cite{Andrade:2008fa} and  rapidity-even directed flow at midrapidity \cite{Gardim:2011qn}, HBT radii \cite{Socolowski:2004hw}, anisotropic flow Fourier coefficients \cite{Gardim:2012yp}, long-range structures observed in two-particle correlations \cite{Takahashi:2009na} and their trigger-angle dependence \cite{Qian:2012qn}.
In this work we use the same simulations that were previously shown to agree with measurements of $v_2$, $v_3$, and $v_4$, across a wide range of centrality and transverse momentum \cite{Gardim:2012yp}. 

Since NeXSPheRIO has thus far shown good agreement with all observables to which it has been compared at RHIC, it provides an ideal baseline prediction for new RHIC measurements.  Any deviation from these predictions will provide valuable and non-trivial information regarding the initial stage of a heavy-ion collision and/or properties of the medium. 

In order to probe what information can be obtained from these measurements and to better understand the underlying physics, we also perform a small number of exploratory calculations.
We  use a number of models to calculate the spatial eccentricities $\varepsilon_n$ of the initial state.  Since it has been shown to a good approximation that $v_2\propto \varepsilon_2$ and $v_3\propto \varepsilon_3$ in each event \cite{Gardim:2011xv,Gardim:2014tya} at least for non-peripheral collisions \cite{Noronha-Hostler:2015dbi}, we can efficiently do high-statistics studies of $v_2$ and $v_3$ correlations.  Specifically, in this work we use the models MC-Glauber, MC-KLN \cite{Drescher:2007ax}, rcBK-MC \cite{Dumitru:2012yr}, and IP-Glasma \cite{Schenke:2012wb}.   

More precisely, if we define the standard eccentricities of the initial entropy density $\rho$ as
\begin{equation}
\varepsilon_n \equiv \left| \frac {\int d^2r\ r^n e^{in\phi} \rho(r,\phi)} {\int d^2r\ r^n  \rho(r,\phi)} \right| ,
\end{equation}
and note the event-by-event relation $v_n \propto \varepsilon_n$ for $n\leq3$, then the following symmetric cumulant should be approximately equal to the normalized ${\rm NSC}(3,2)$ after hydrodynamic evolution:
\begin{align}
\label{epsSC}
{\rm \varepsilon SC}(3,2) &\equiv  \frac {\langle \varepsilon_3^2 \varepsilon_2^2 \rangle -  \langle \varepsilon_3^2\rangle \langle \varepsilon_2^2 \rangle} {\langle \varepsilon_3^2\rangle \langle \varepsilon_2^2 \rangle} .
\end{align}

\begin{figure}
\includegraphics[width=\linewidth]{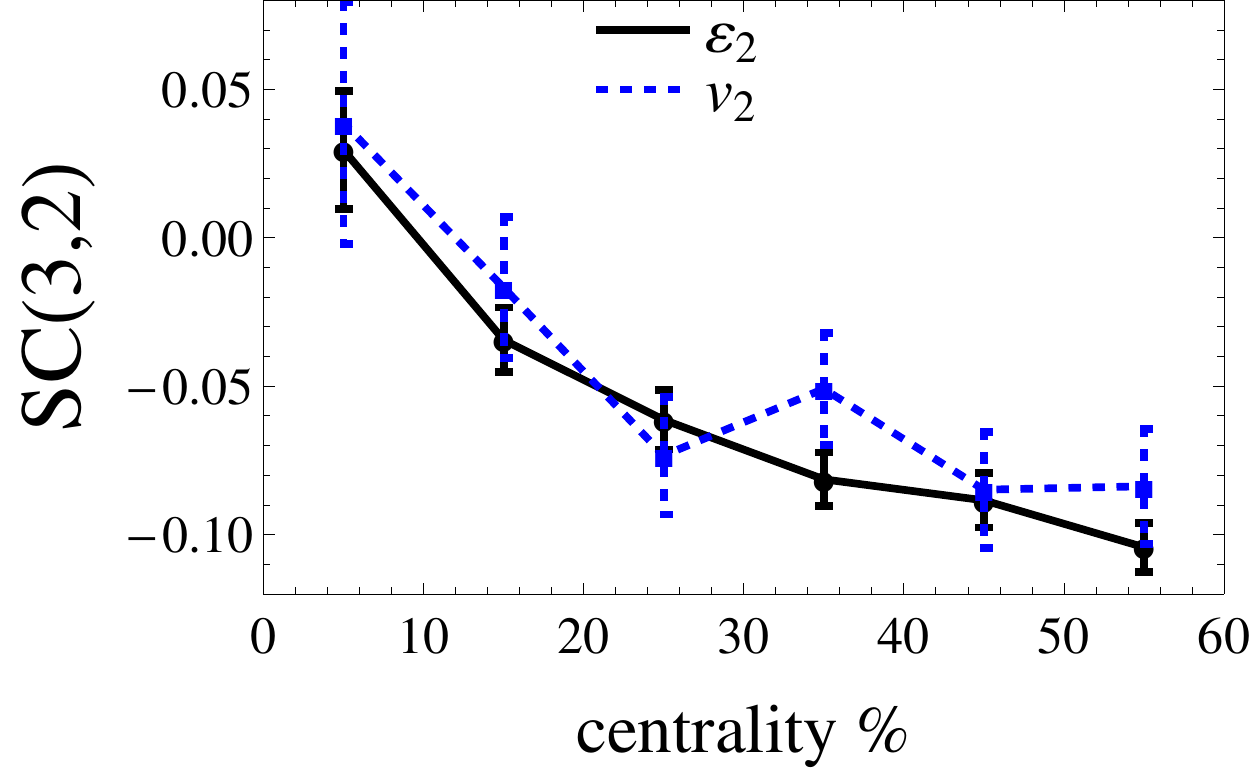}
\caption{\label{vnen} $\varepsilon{\rm SC}(3,2)$ \eqref{epsSC} and ${\rm NSC}(3,2)$ \eqref{NSC} from NeXSPheRIO. Points are shifted horizontally for readability.  Note that neither centrality rebinning nor non-trivial event weights were used here, as discussed in the following section, and as used in our final prediction.  Errors are statistical, obtained via jackknife resampling.}
\end{figure}

In Fig.~\ref{vnen} we show explicitly that there is a small, if any, difference between $\varepsilon{\rm SC}(3,2)$ and ${\rm NSC}(3,2)$, using our NeXSPheRIO results.
Thus, we can study the dependence of NSC(3,2) on various factors without need to run computationally-intensive hydrodynamic simulations.   This allows us to generate more statistics, as well as to vary models and parameters.

For the MC-Glauber model, we also evolve the initial conditions through a 2+1 D viscous hydrodynamic code, to study $v_n$ for $n>3$ and the effects of viscosity.  We do the same for MC-KLN initial conditions at LHC energy.  For details, see Ref.~\cite{Noronha-Hostler:2013gga,Noronha-Hostler:2014dqa,Noronha-Hostler:2015coa}

These ancillary results can be found in Appendix \ref{anc}.
\section{Results and Conclusions}
We begin by noting that details of the experimental analysis can have a significant effect on measurements and calculations.  

\begin{figure}
\includegraphics[width=\linewidth]{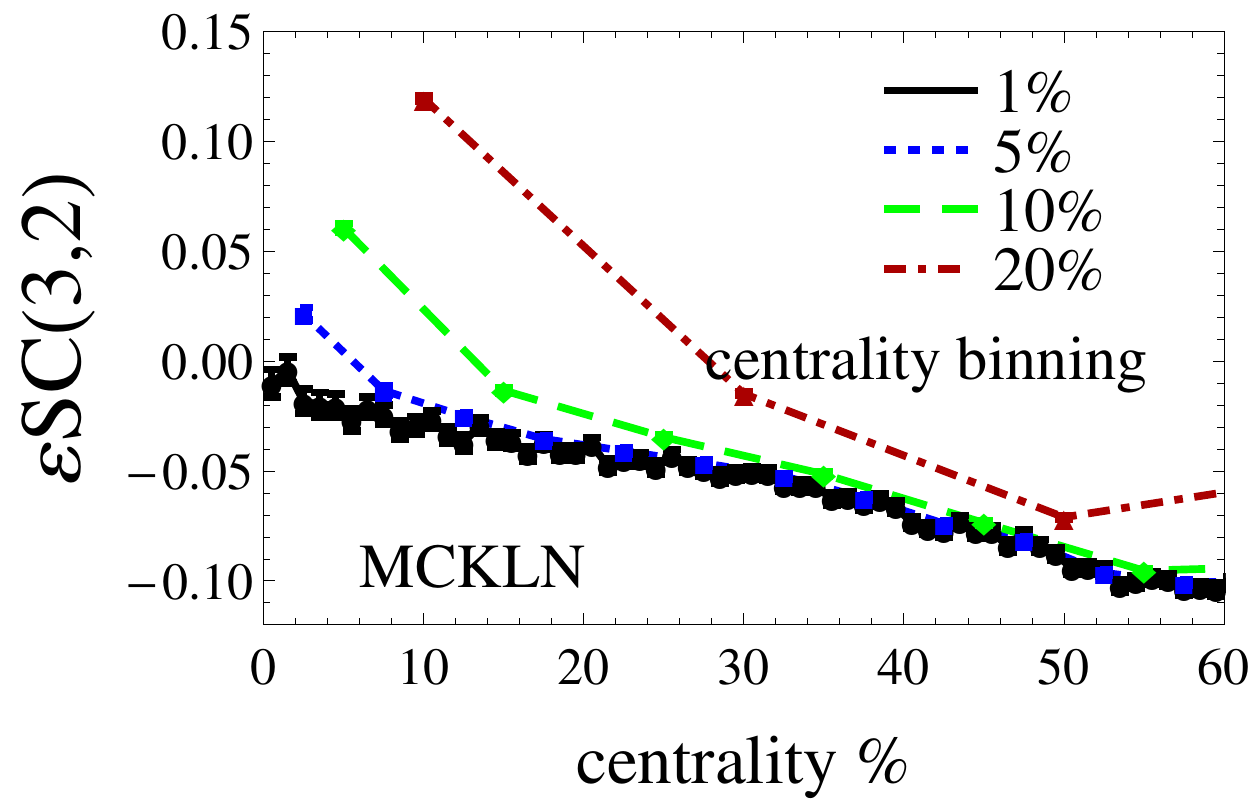}
\caption{\label{bins} $\varepsilon{\rm SC}(3,2)$ from the MC-KLN model \cite{Drescher:2007ax}, calculated with various centrality bin sizes.  Large centrality bins have a positive contribution from intra-bin centrality fluctuations and the monotonic dependence of $v_2$ and $v_3$ on centrality. Events are weighted equally.  Errors are statistical, obtained via jackknife resampling (only visible in the results for 1\% centrality bins).}
\end{figure}

One detail is the centrality binning, which can change measurements of symmetric cumulants.  This is mainly due to the following effect:  on average, more peripheral collisions have larger $v_n$ for all $n$, while more central collisions have smaller $v_n$.   If one measures ${\rm NSC}(n,m)$ using events in a large range of centrality, the impact parameter will fluctuate significantly within the bin.  This trivial effect will generate a spurious positive correlation, compared to the value obtained when using narrow centrality bins.  Since the dependence of $v_n$ on centrality is strongest in central collisions, this effect is most important there.  We illustrate this in Fig.~\ref{bins}, where $\varepsilon{\rm SC}(3,2)$ is calculated using various bin sizes. (Note that in Fig.~\ref{bins} each event has equal weight as opposed to multiplicity-dependent weights Eq.~(\ref{w24}), as discussed below.)  

Since this effect does not represent interesting unknown physics, it is preferable to use small centrality bins.  This may be a particular concern for RHIC measurements, since limited statistics will likely demand large centrality bins.  However, one can always do the analysis in smaller bins, and then recombine them to improve statistics. 

In all of the following, we perform the analysis in 1\% centrality bins, which are then recombined into 10\% bins whenever necessary for reducing statistical uncertainty.

\begin{figure}
\includegraphics[width=\linewidth]{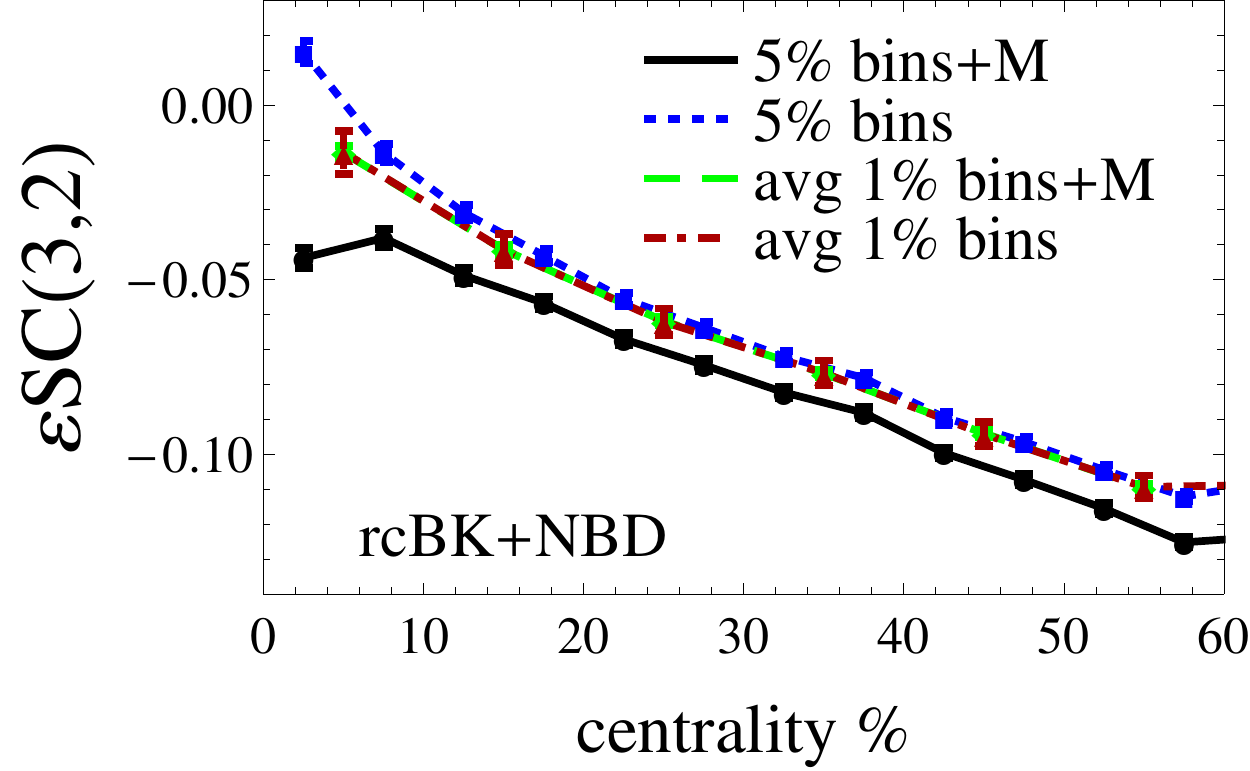}
\caption{\label{fig:wt} $\varepsilon{\rm SC}(3,2)$ from the rcBK-MC model \cite{Drescher:2007ax}, calculated with and without event weights, Eqs.~\eqref{w24} (where M is taken to be the gluon multiplicity), and with both 1\% and 5\% centrality bin widths.  The 1\% bin results have been recombined into 10\% windows to reduce statistical errors, which have been calculated via jackknife resampling.}
\end{figure}

The second important detail of the experimental analyses that have been done of the symmetric cumulants NSC($n,m$) \cite{ALICE:2016kpq} is the non-unity event-weights used in the analysis [see Eqs.~\eqref{weight} and \eqref{w24}].  The addition of this multiplicity-dependent weight when taking the event average has a non-negligible effect, as shown in Fig.~\ref{fig:wt}.  Since the multiplicity weighing of the 4 particle correlation is different than that of the 2 particle correlation in Eq.~\eqref{NSC}, the former term is biased toward a larger multiplicity and, therefore, a lower $v_n$.  As a result, measurements in larger centrality bins now are \textit{lower} than those with smaller centrality bins.   Thus,  calculations that want to make direct comparison to experimental data must take multiplicity weighing into account.  Just as in the case with unit weights, small bins are preferable, to reduce this bias.   When using small centrality bins, the weighting scheme has a negligible effect (see the 1\% curves in Fig.~\ref{fig:wt}).  In all of the following we use the same weights the ALICE collaboration used \eqref{w24}, in addition to 1\% centrality binning that is reaveraged into 10\% bins to improve statistics.

Such a recombination can involve a simple average, or one can give a non-trivial weight to each sub-bin.  A common choice for 4-particle cumulants is to use the average of $W_{\langle 4\rangle}$ as the weight of each sub-bin (see Eq.~\eqref{w24}), which we also use in all of the following.  Explicitly,

\begin{eqnarray}
{\rm NSC}_{10\%}&=&\frac{\sum_{c=1}^{10} {\rm NSC}_{1\%} \sum_{\rm events} W_{\langle 4\rangle} }{\sum_{c=1}^{10} \sum_{\rm events}  W_{\langle 4\rangle}} ,
\end{eqnarray}
where the first sum is over the 1\% centrality sub-bins, labelled by $c$, and the second sum is over events within a sub-bin.

Note that experiments typically select centrality in a different region of pseudorapidity from the main measurement that defines the event's multiplicity. 
Because of this, multiplicity fluctuations in these measurements are even larger than in our calculation, in which the same multiplicity is used for centrality selection and for weighting, and so the effect may be even larger in experiment.

A discussion of these effects in existing LHC measurements and the comparison to calculations can be found in Appendix \ref{anc}.

\begin{figure}
\includegraphics[width=\linewidth]{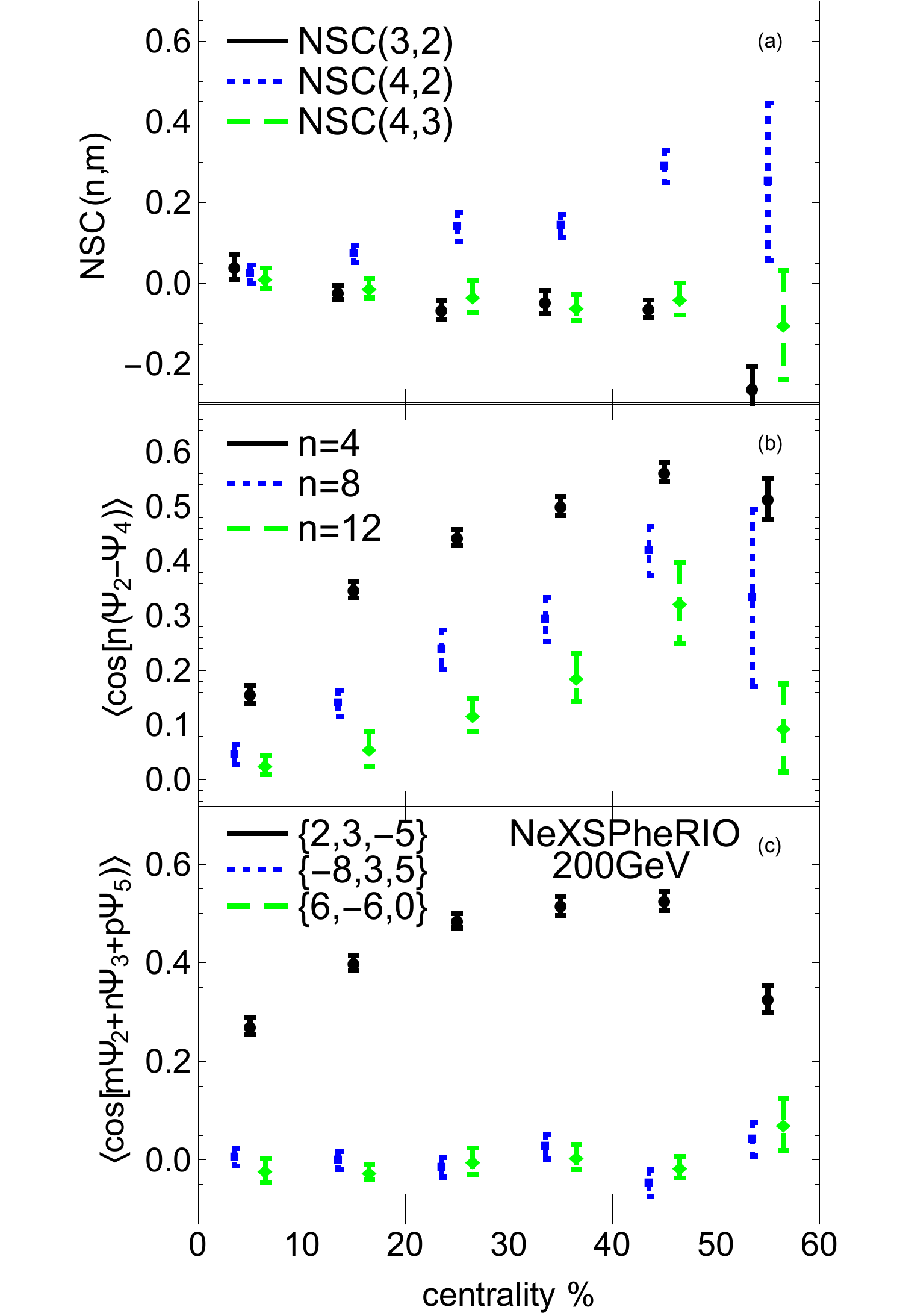}
\caption{\label{pred} Predictions for NSC$(n,m)$ \eqref{NSC} and mixed harmonic correlations \eqref{EP24}, \eqref{EP23}, \eqref{EP235}, in 200 GeV Au+Au collisions from NeXSPheRIO for $p_T>200$ MeV and $|\eta| < 1$.  See text for details.  Points are shifted horizontally for readability.  Errors are statistical, obtained via jackknife resampling.}
\end{figure}

Our final predictions for mixed-harmonic correlations in 200 GeV Au-Au collisions, with STAR kinematic cuts, are presented in Fig.~\ref{pred}.
We remark that the general size and centrality dependence of each observable are similar to available measurements at the LHC \cite{Aad:2014fla, ALICE:2016kpq} (though typically the RHIC magnitude is smaller).  As such, we do not expect any drastic change in the RHIC measurement.  

Nevertheless, a precise quantitative comparison to measurements at RHIC will provide needed guidance to discriminate between different theoretical models.
\acknowledgments
We thank M. Nasim, A. Timmins, Y. Zhou, and A. Bilandzic for discussion and information about the experimental analyses at LHC and RHIC.
J.N.H.~was supported by the National Science Foundation under grant no.~PHY-1513864. F.G.~acknowledges support from Funda\c{c}\~ao de Amparo \`a Pesquisa do Estado de S\~ao Paulo (FAPESP grants 2015/00011-8 and 2016/03274-2) and Conselho Nacional de Desenvolvimento Cient\'{\i}fico e Tecnol\'ogico (CNPq grant 310737/2013-3).  F.G.G.~was supported by Conselho Nacional de Desenvolvimento Cient´ıfico e Tecnol´ogico (CNPq) No. 449694/2014-3, and FAPEMIG.
\appendix
\section{Dependence of NSC($3,2$) on viscosity, collision energy, and initial conditions, and effect of centrality binning on LHC data}
\label{anc}
\begin{figure}
\includegraphics[width=\linewidth]{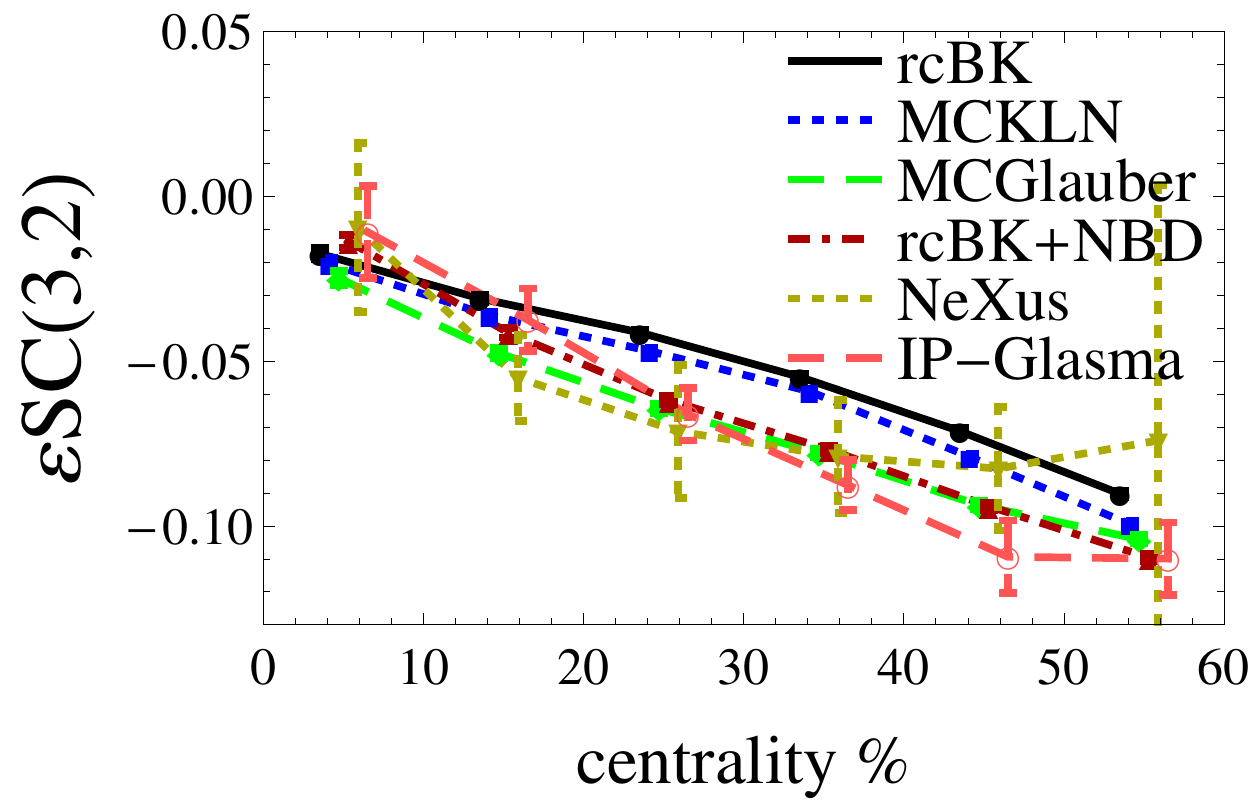}
\caption{\label{eccs} $\varepsilon{\rm SC}(3,2)$ from various models for hydrodynamic initial conditions:  MC-Glauber, MC-KLN \cite{Drescher:2007ax}, rcBK-MC with and without negative binomial fluctuations \cite{Dumitru:2012yr}, and IP-Glasma \cite{Schenke:2012wb}. The cumulant is calculated in 1\% bins, which are recombined into 10\% bins, to eliminate spurious correlation. See text for details.   Points are shifted horizontally for readability.   Errors are statistical, obtained via jackknife resampling.}
\end{figure}

In Fig.~\ref{eccs}, we show a calculation of $\varepsilon{\rm SC}(3,2)$ from various models for initial conditions.   The normalized symmetric cumulant does not appear to vary by a large amount, despite the fact that each of these models is quite different.  

\begin{figure}
\includegraphics[width=\linewidth]{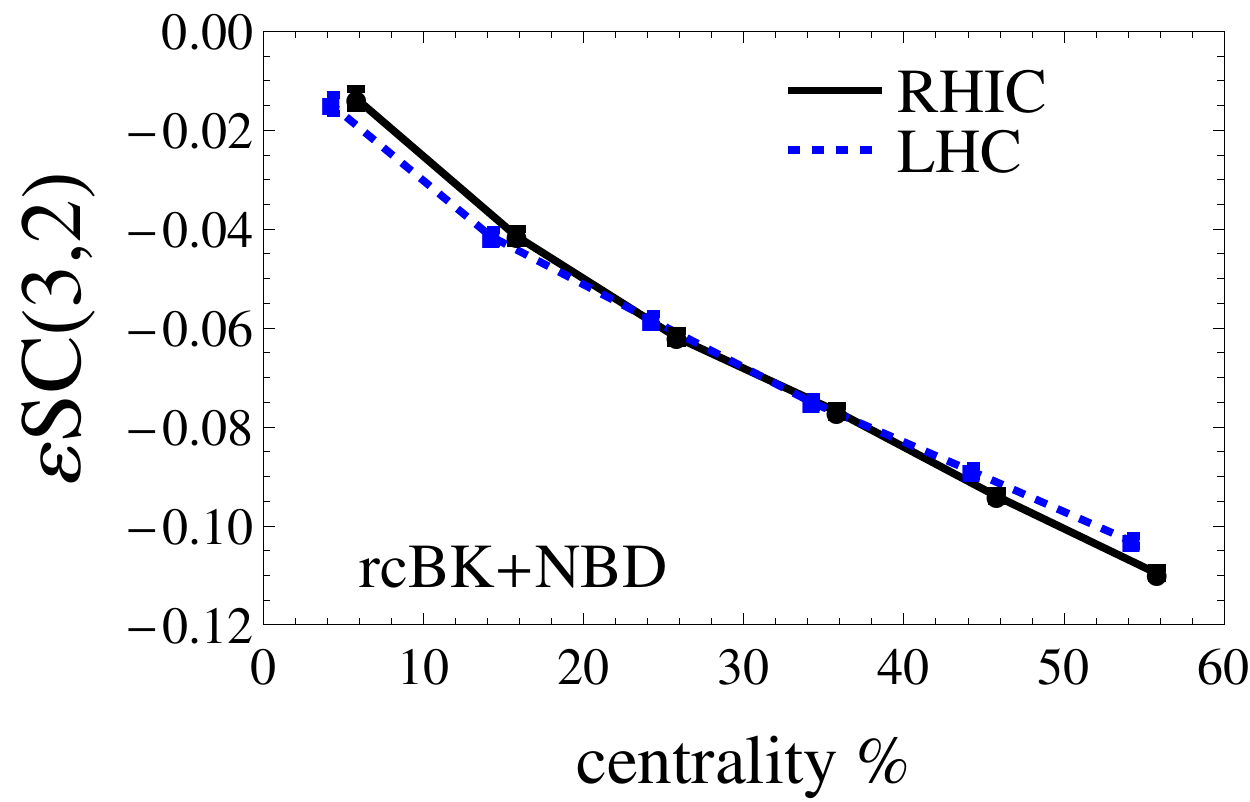}
\caption{\label{en} $\varepsilon{\rm SC}(3,2)$ from in the MC-KLN model for 200 GeV Au-Au collisions compared to 2.76 TeV Pb-Pb collisions.  Points are shifted horizontally for readability.   Errors are statistical, obtained via jackknife resampling.}
\end{figure}

The dependence of NSC(3,2) on collision energy is very small, as illustrated in Fig.~\ref{en} for MC-KLN initial conditions.    We have verified that this is true for all of the models considered here.

\begin{figure}
\includegraphics[width=\linewidth]{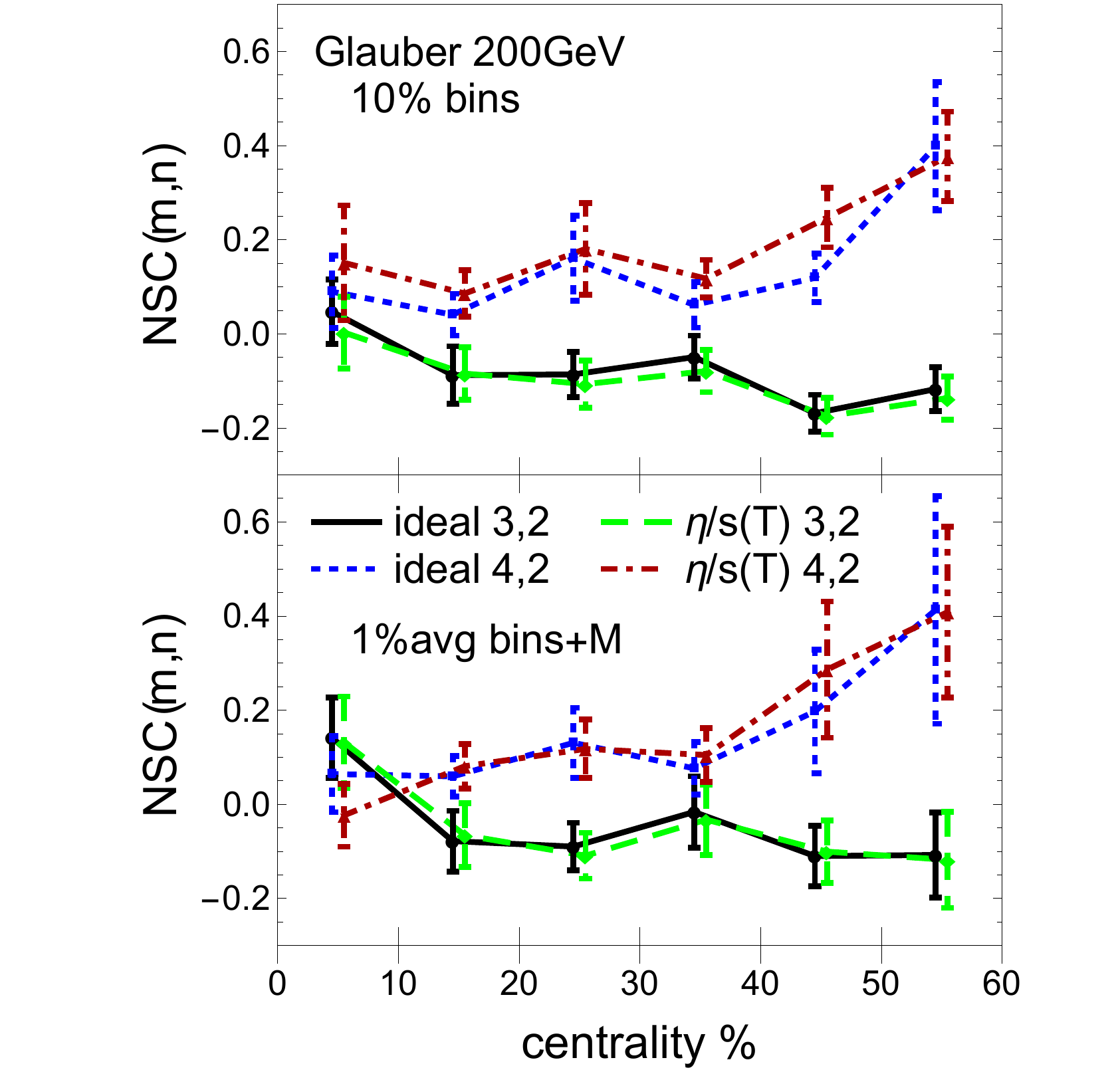}
\caption{\label{visc} ${\rm NSC}(n,m)$ from hydrodynamic calculations using Glauber initial conditions and zero shear viscosity, compared to the temperature-dependent shear viscosity from Ref.~\cite{NoronhaHostler:2012ug,NoronhaHostler:2008ju,Nakamura:2004sy}. The results in the upper panel are calculated with 10\% centrality bins and unit event weights, while the bottom panel uses multiplicity weights and 1\% recombined centrality bins. Points are shifted horizontally for readability.   Errors are statistical, obtained via jackknife resampling.}
\end{figure}

In Fig.~\ref{visc}, we show calculations of NSC($3,2$) and NSC($4,2$) using MC-Glauber initial conditions, with zero shear viscosity and a nonzero temperature-dependent shear viscosity (from Ref.~\cite{NoronhaHostler:2012ug,NoronhaHostler:2008ju,Nakamura:2004sy}).  In general, the dependence on viscosity is not large.  

Note that the same events are used for the ideal and viscous calculations.  As such, the statistical errors are strongly correlated.  We have checked that there is statistically significant dependence of NSC(4,2) on viscosity, but not NSC(3,2), when one does the calculation in a typical way with wide (10\%) centrality bins, and equal event weighting.   Viscosity tends to increase the magnitude of the correlation between $v_2^2$ and $v_4^2$. 

However, using multiplicity weights and 1\% recombined centrality bins removes any statistically-significant dependence of NSC(4,2) on viscosity.    Therefore, even investigations of viscosity dependence can be affected by analysis details, which should be taken into account.

We note in particular that NSC(3,2) seems to depend little on the model of initial conditions, viscosity, and collision energy.  This makes it a very interesting observable, as a robust test of our current hydrodynamic paradigm, and which is independent of what we believe to be our most important uncertainties.

Further study of the dependence on viscosity and initial conditions can now found in Ref.~\cite{Zhu:2016puf}.

\begin{figure}
\includegraphics[width=\linewidth]{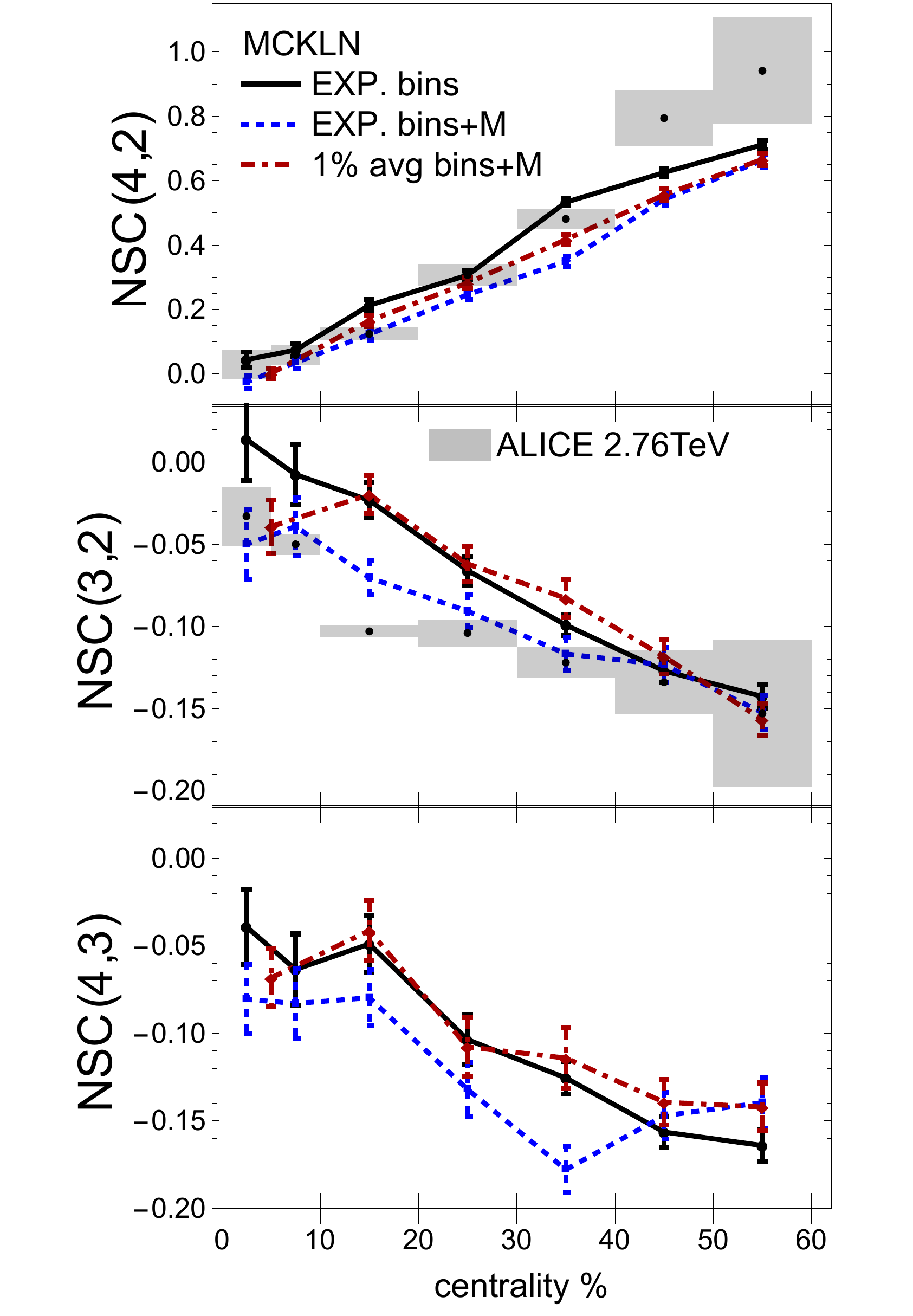}
\caption{\label{LHC} ${\rm NSC}(n,m)$ for 2.76 GeV Pb-Pb collisions from hydrodynamic calculations using MC-KLN initial conditions, with two different centrality binning schemes, with and without multiplicity weights \eqref{w24}, compared to measurements from the ALICE Collaboration \cite{ALICE:2016kpq}. 
Points are shifted horizontally for readability.   
Errors are statistical, obtained via jackknife resampling.}
\end{figure}

Next, we note the effect of centrality binning (plus multiplicity-dependent event weighting) on the NSC measurements done by ALICE.  In Fig.~\ref{LHC}, we show a hydro calculation of NSC(3,2) and NSC(4,2) with the analysis done the same way as ALICE (a mix of 5\% and 10\% centrality bins with multiplicity weights) \cite{ALICE:2016kpq}, compared to an improved analysis with 1\% bins, recombined into 10\% windows, as well as a calculation with the wider ALICE binning but without multiplicity weights.  
We note that agreement with NSC(3,2) data is significantly improved when the experimental analysis is mimicked, and the switching of binning explains some of the seeming discontinuity at 10\% centrality.  Agreement with NSC(4,2), on the other hand, appears to get worse with the corrected analysis, at least in this calculation.  In any case, it is clear that any precise comparison to the ALICE data must take into account the correct event weighting.

\begin{figure}
\includegraphics[width=\linewidth]{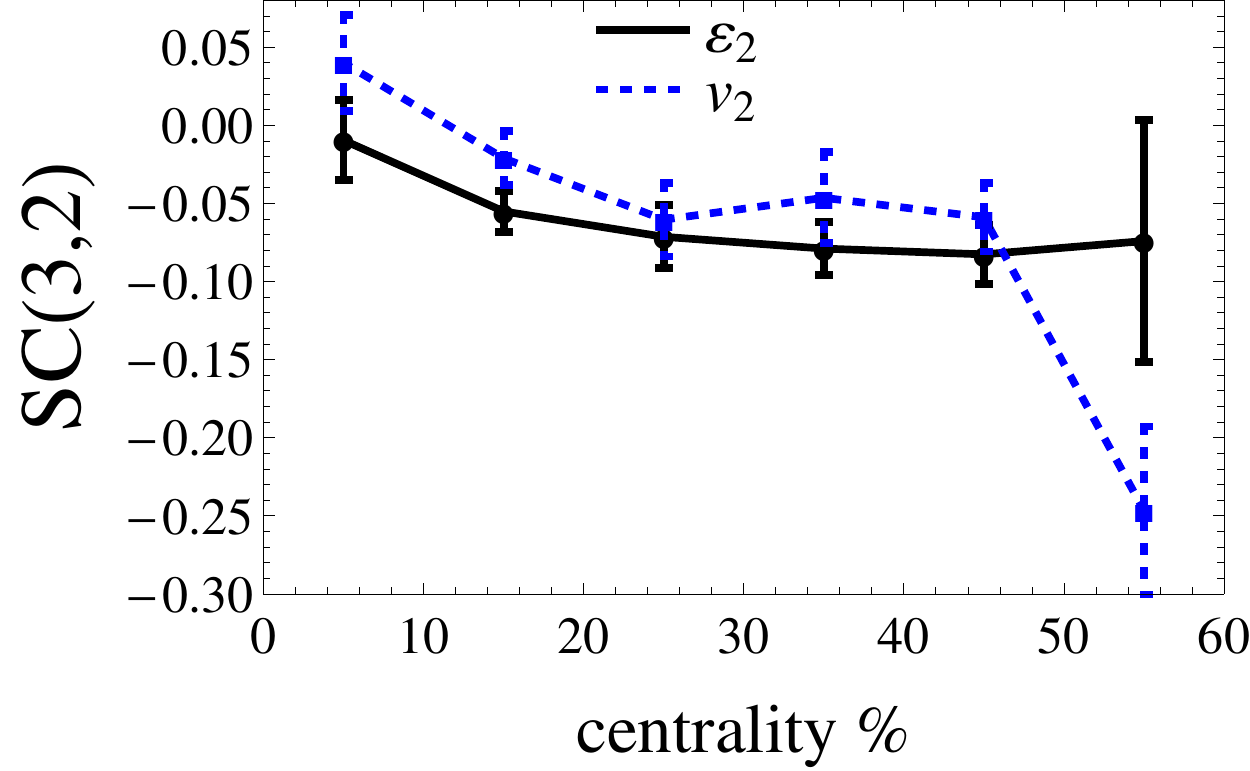}
\caption{\label{vnenExp} $\varepsilon{\rm SC}(3,2)$ \eqref{epsSC} and ${\rm NSC}(3,2)$ \eqref{NSC} from NeXSPheRIO.  As Fig.~\ref{vnen} except the analysis was done with 1\% centrality bins and multiplicity weights Eq.~\eqref{w24}.  Points are shifted horizontally for readability.  Errors are statistical, obtained via jackknife resampling.}
\end{figure}

Finally, in Fig.~\ref{vnenExp}, we repeat Fig.~\ref{vnen}, except using multiplicity weights and 1\% centrality bins.  In this case, there is a statistically-significant difference between NSC(3,2) and $\varepsilon{\rm SC}(3,2)$, showing again that analysis details can be important, and illustrating that the proportionality between $v_n$ and $\varepsilon_n$ is not perfect.

\end{document}